%% file: author/0_main.tex
\documentclass[graybox]{svmult}

\usepackage{physics}
\usepackage{type1cm}        
\usepackage{makeidx}         
\usepackage{graphicx}        
\usepackage{multicol}        
\usepackage[bottom]{footmisc}

\usepackage{newtxtext}       %
\usepackage[varvw]{newtxmath}       
\usepackage{soul}
\usepackage{tabularx}
\usepackage{array}
\usepackage{enumitem}

\makeindex             

\begin{document}

\title*{Adversarial Threats in Quantum Machine Learning: A Survey of Attacks and Defenses}
\author{Archisman Ghosh\orcidID{0000-0002-0264-6687}, Satwik Kundu\orcidID{0000-0002-2140-6486} and Swaroop Ghosh\orcidID{0000-0001-8753-490X}}
\institute{Archisman Ghosh \at Pennsylvania State University, University Park, PA, USA \email{apg6127@psu.edu}
\and Satwik Kundu \at Pennsylvania State University, University Park, PA, USA \email{satwik@psu.edu}
\and Swaroop Ghosh \at Pennsylvania State University, University Park, PA, USA \email{szg212@psu.edu}}
%
%
\maketitle


\abstract{Quantum Machine Learning (QML) integrates quantum computing with classical machine learning, primarily to solve classification, regression and generative tasks. However, its rapid development raises critical security challenges in the Noisy Intermediate-Scale Quantum (NISQ) era. This chapter examines adversarial threats unique to QML systems, focusing on vulnerabilities in cloud-based deployments, hybrid architectures, and quantum generative models. Key attack vectors include model stealing via transpilation or output extraction, data poisoning through quantum-specific perturbations, reverse engineering of proprietary variational quantum circuits, and backdoor attacks. Adversaries exploit noise-prone quantum hardware and insufficiently secured QML-as-a-Service (QMLaaS) workflows to compromise model integrity, ownership, and functionality. Defense mechanisms leverage quantum properties to counter these threats. Noise signatures from training hardware act as non-invasive watermarks, while hardware-aware obfuscation techniques and ensemble strategies disrupt cloning attempts. Emerging solutions also adapt classical adversarial training and differential privacy to quantum settings, addressing vulnerabilities in quantum neural networks and generative architectures. However, securing QML requires addressing open challenges such as balancing noise levels for reliability and security, mitigating cross-platform attacks, and developing quantum-classical trust frameworks. This chapter summarizes recent advances in attacks and defenses, offering a roadmap for researchers and practitioners to build robust, trustworthy QML systems resilient to evolving adversarial landscapes.}

\input{author/section1}
\input{author/section2}

\input{author/section3}
\input{author/section4}

\begin{acknowledgement}
The work is supported in parts by NSF (CNS-2129675, CCF-2210963), gifts from Intel and IBM Quantum Credits.
\end{acknowledgement}
\ethics{Competing Interests}{
The authors have no conflicts of interest to declare that are relevant to the content of this chapter.}


\bibliographystyle{IEEEtran} 
\bibliography{author/refs}
\end{document}

%% file: author/section1.tex
\section{Introduction}
\label{sec:1}

As QML rapidly matures from theoretical promise to experimental and early-stage practical deployment, concerns around its robustness and security have gained significant urgency \cite{Biamonte2017, Schuld03042015, Cerezo2022}. In the current NISQ era \cite{Preskill2018quantumcomputingin}, QML systems are inherently constrained by hardware limitations, noisy gate operations, and limited qubit connectivity. These physical limitations are further complicated by the architectural trends that dominate QML today: hybrid classical-quantum designs, cloud-based quantum computing access \cite {nguyen2024quantumcloudcomputingreview}, and variational quantum circuits (VQCs) \cite{farhi2018classificationquantumneuralnetworks} trained via classical optimizers.
While these developments make QML feasible and scalable in the near term, they also introduce a diverse and novel attack surface. Adversarial threats that target classical ML pipelines do not transfer trivially into the quantum domain, but instead manifest in new forms due to the fundamentally different representations, computational models, and physical substrates involved. Crucially, QML systems inherit and amplify certain vulnerabilities, such as model theft, data poisoning, and integrity breaches, particularly when deployed via QML-as-a-Service (QMLaaS) platforms or accessed through remote execution interfaces \cite{kundu2024securityconcernsquantummachine}.
This chapter is motivated by the urgent need to understand, categorize, and mitigate adversarial threats specific to QML. As with any emerging computational paradigm, security considerations must evolve in parallel with functionality. QML is not exempt from this imperative.

\subsection{Vulnerabilities in the execution flow of QML}

\begin{figure}[t]
    \centering
    \includegraphics[scale=.55]{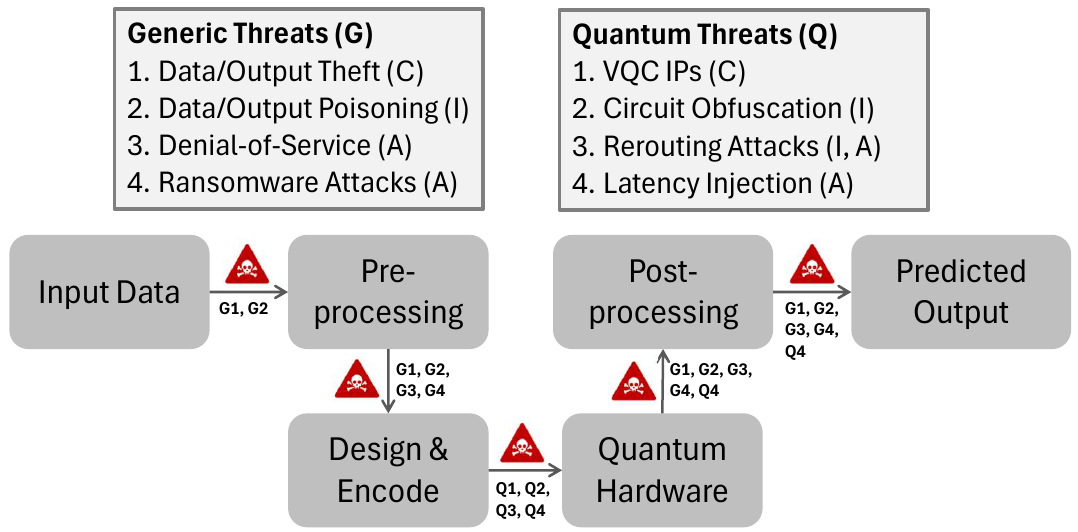}
    \caption{Key Threats to Confidentiality (C), Integrity (I), and Availability (A) in the QML Pipeline.}
    \label{fig:threats}       
\end{figure}

QML enables advanced data-driven inference and decision-making using machine learning paradigms in the quantum Hilbert space. While promising, this integration introduces distinct security challenges stemming from the inherent features of quantum systems, such as superposition and entanglement, and the practical limitations of NISQ devices. These factors expose QML pipelines to a range of adversarial threats, necessitating the development of tailored security mechanisms (Fig. \ref{fig:threats}).
The QML workflow begins with preprocessing and encoding classical data into quantum states, often via schemes like amplitude or angle encoding \cite{lloyd2020quantumembeddingsmachinelearning, Schuld_2019, Havl_ek_2019}. This stage is vulnerable to inference attacks, where adversaries with access to circuits or their transpiled forms can deduce the data encoding methods, compromising input confidentiality. Additionally, data poisoning attacks can be mounted during this phase by injecting malicious perturbations that disrupt training dynamics and degrade model reliability \cite{ghosh2024quantumimitationgamereverse}.
Following data encoding, the construction and initialization of parameterized quantum circuits (PQCs) form the computational core of QML models. These circuits are susceptible to model extraction, wherein unauthorized access leads to replication or approximation of proprietary designs. Moreover, adversaries may introduce stealthy backdoors, malicious circuit modifications that trigger incorrect outputs under specific conditions, posing risks to the integrity and trustworthiness of the deployed models \cite{upadhyay2023stealthyswapsadversarialswap}.

QML models, often, adopt a hybrid architecture, wherein quantum circuits are evaluated by quantum processors while classical optimizers iteratively update circuit parameters. This interplay between quantum and classical components introduces unique security risks. Data exchanged during this loop, such as intermediate outputs, gradients, or loss values, may be intercepted, enabling adversaries to infer sensitive model parameters or internal structures \cite{xiong2020improvedadversarialtraininglearned}. Furthermore, manipulation of the classical optimization process itself can misguide the training trajectory, resulting in degraded performance or the embedding of malicious behavior within the model \cite{sofer2025unveilingmitigatingadversarialvulnerabilities}.
 The deployment setup of QML models on cloud-based platforms, while scalable, exposes the system to inference-time threats. Publicly accessible APIs allow adversaries to probe the model and perform functionality reconstruction or model extraction attacks. Additionally, side-channel vulnerabilities may arise from quantum hardware characteristics, such as execution timing, thermal output, or noise signatures, that can inadvertently leak sensitive information about the model’s structure or computations \cite{choudhury2024crosstalkinducedchannelthreatsmultitenant, tan2025qubithammerattacksqubitflipping}.

These multifaceted vulnerabilities across the QML execution flow highlight the urgent need for holistic, quantum-aware security strategies. Safeguarding QML requires addressing both classical and quantum threats to ensure robust and trustworthy model deployment.

\subsection{Why is theft of QML models a concern?}
The vulnerabilities outlined across the QML execution pipeline are not merely technical limitations, they represent tangible risks with significant implications. The theft or manipulation of QML models can compromise the confidentiality of sensitive input data, erode trust in quantum-enabled inference, and undermine the proprietary value of carefully engineered quantum circuits. Critically, both quantum data and model training incur substantial costs: preparing high-quality quantum datasets often involves complex preprocessing and domain-specific encoding strategies \cite{ghosh2024guardiansquantumgan}, while training variational quantum circuits demands extensive quantum-classical iteration, subject to hardware constraints and limited quantum coherence times \cite{kundu2024evaluating}. These investments make QML systems highly valuable assets and, consequently, attractive targets for adversaries. As deployment increasingly shifts to cloud-hosted platforms and QMLaaS offerings, the incentives for unauthorized access, replication, or sabotage are amplified. The inherent opacity of quantum operations further complicates detection and attribution, making attacks stealthier and potentially more damaging.

%% file: author/section2.tex
\section{Attack models}
\label{sec2}

Developing and applying well-defined attack models is essential for advancing the study of adversarial robustness in QML. Given the unique architectural, algorithmic, and physical characteristics of QML systems, security research in this space requires a clear understanding of adversarial assumptions, access levels, and objectives. Carefully constructed attack models enable researchers to map existing vulnerabilities, identify overlooked threat vectors, and avoid proposing unrealistic or impractical defenses. As with other domains of security research, such as hardware Trojans or classical adversarial ML, choosing the appropriate attack model is a foundational step. It not only informs the design of effective countermeasures but also guides reproducibility, benchmarking, and comparison across studies. In what follows, we describe a set of representative attack models for QML that serve to categorize current adversarial efforts, surface emerging trends, and offer a framework for evaluating the security posture of QML architectures across deployment contexts.

\begin{table}[ht]
\centering
\caption{Taxonomy of Quantum Adversarial Attack Models}
\begin{tabularx}{\textwidth}{|X|X|X|X|X|}
\hline
\textbf{Threat Model} & \textbf{Attacker Profile} & \textbf{Target Artifacts} & \textbf{Capabilities} & \textbf{Related Works} \\
\hline
Black-box & External user / API client; No Internal access & QML outputs & Model stealing, functionality replication & \cite{kundu2024evaluating, fu2025copyqnnquantumneuralnetwork, choudhury2024crosstalkinducedchannelthreatsmultitenant, xu2024securityattacksabusingpulselevel} \\
\hline
Gray-box & Cloud provider / semi-privileged adversary; partial access  & Transpiled circuits & Manipulate inputs or IRs, estimate parameters or logic & \cite{kundu2025adversarialdatapoisoningattacks, wang2023qumosframeworkpreservingsecurity} \\
\hline
White-box & Transpiler/ infrastructure insider; full access to model internals& PQC, Transpiled gates, pulse schedules & Analyze and RE circuits, inject Trojans & \cite{ghosh2024quantumimitationgamereverse, ghosh2024aidrivenreverseengineeringqml, xu2024securityattacksabusingpulselevel, 10.1145/3358184, rehman2025opaqueobfuscatingphasequantum}\\
\hline
\end{tabularx}
\label{tab:attack-taxonomy}
\end{table}

\subsection{Taxonomy of the attack models}
Attack models in QML can be broadly categorized according to the adversary’s level of access to the model and execution environment (Table \ref{tab:attack-taxonomy}). These categories--black-box, gray-box, and white-box--provide a foundational framework for assessing the feasibility, impact, and countermeasures for a range of threat vectors.

\begin{figure}[t]
    \centering
    \includegraphics[scale=.58]{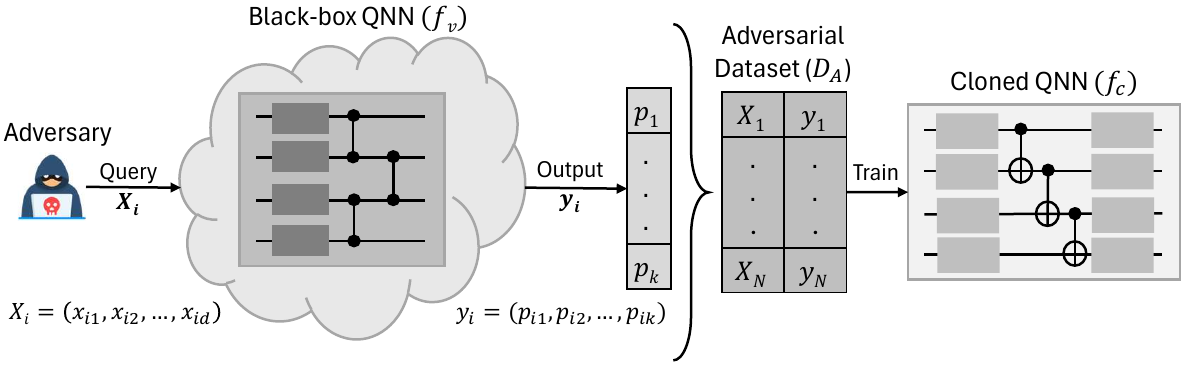}
    \caption{An adversary sends a query vector $X_i = (x_{i1}, x_{i2}, ..., x_{id})$ to a cloud-based victim QNN ($f_v$), receiving a vector of class probabilities ($f_v(X_i) = y_i = (p_{i1}, p_{i2}, ..., p_{ik})$) in response. The adversary repeats this to build an attacker dataset $D_A$ and trains a substitute model $f_c$ to clone $f_v$'s functionality.}
    \label{fig:model-extraction}       
\end{figure}

\subsubsection{Black-Box Attack Models}
In the black-box setting, adversaries do not possess internal knowledge of the QML model architecture, circuit parameters, or training data. Access is limited to querying the model via an external interface, common in QMLaaS or cloud-based deployments. Even with these constraints, attackers can mount effective threats.

\begin{itemize}[label={--}]

    \item \textbf{Model Extraction and Counterfeit Generation:} A commonly encountered threat in QMLaaS deployments is the adversary modeled as an external user who lacks access to the internal architecture, parameterization, training data, or compilation artifacts of the target Quantum Neural Networks (QNNs) \cite{kundu2024evaluating} (Fig. \ref{fig:model-extraction}). The attacker interacts with the model solely through a public inference API, submitting classical inputs and receiving output distributions, typically Top-1 predictions or full class probability vectors (Top-$k$). Despite the lack of introspective access, the adversary can mount a model extraction attack by systematically querying the QNN with a large corpus of input samples and collecting the corresponding outputs to construct a surrogate dataset that approximates the decision boundary of the original model. To improve fidelity in the presence of NISQ-induced noise, the adversary may issue repeated queries across different time intervals and apply a variance-based consistency check to filter unreliable outputs. This denoised dataset is then used to train a counterfeit QNN locally. Advanced techniques such as quantum-domain contrastive learning are employed to pre-train a feature encoder on auxiliary datasets, while quantum transfer learning is used to fine-tune a classifier based on the cleaned query outputs. The attacker’s objective is to construct a functionally equivalent substitute model that replicates the behavior of the original QNN with high fidelity, thereby bypassing service access controls, evading usage metering, and undermining the confidentiality and economic value of the proprietary model. This demonstrates that even under restricted black-box conditions, it is feasible to exfiltrate the functional essence of a QNN through strategic input-output interaction alone \cite{fu2025copyqnnquantumneuralnetwork}.

    \item \textbf{Crosstalk-Induced Side-Channel Attacks:} In the current NISQ era of multi-tenant quantum computing environment, any realistic and minimally privileged user can pose as an adversarial threat operating within a cloud-based Quantum-as-a-Service (QaaS) platform \cite{choudhury2024crosstalkinducedchannelthreatsmultitenant} \cite{xu2024securityattacksabusingpulselevel}. The attacker executes their quantum circuit on a disjoint subset of qubits concurrently with a victim user, both sharing the same physical quantum processor. Crucially, the attacker does not require co-location of qubits, elevated access, or insider control over the quantum hardware. Instead, the attack leverages crosstalk, a form of physical interference inherent in NISQ devices, between qubits to create a passive side channel. By strategically positioning idle ``snooping" qubits near victim qubits, the attacker detects subtle perturbations induced by two-qubit operations (e.g., $CNOT$ gates) in the victim’s circuit. These perturbations manifest as measurable deviations in the idle qubit states, allowing the adversary to infer temporal and spatial gate activity. With access only to the shared quantum execution timeline and the ability to repeatedly submit crafted sensing circuits, the adversary systematically extracts circuit-level structural features, such as the number and timing of entangling operations, thereby enabling partial or full reconstruction of the victim’s quantum circuit, underscoring that physical-layer interactions can be exploited to compromise user privacy without violating system access controls.

\end{itemize}

\begin{figure}[t]
    \centering
    \includegraphics[scale=.8]{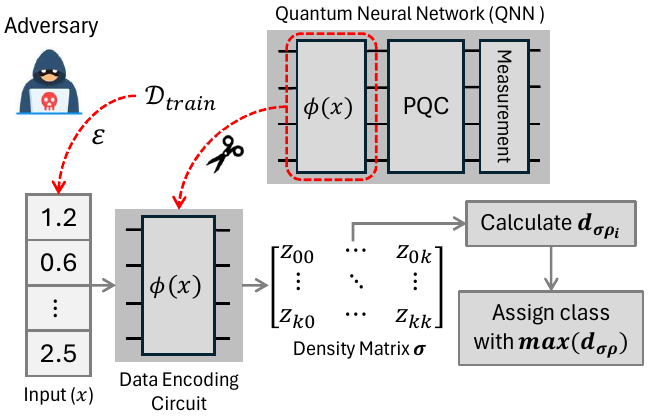}
    \caption{Overview of QUID's label-poisoning technique. Adversary extracts the encoding circuit from the QNN and uses it to compute the output density matrix ($\sigma$) for a portion ($\varepsilon$) of the training data $\mathcal{D}_{train}$. It then calculates the matrix distance between $\sigma$ and the remaining samples in the dataset ($\rho_i$), assigning the class with the max. distance ($\max d_{\sigma\rho}$).}
    \label{fig:poisoning}       
\end{figure}

\subsubsection{Gray-Box Attack Models}
Gray-box attackers possess partial knowledge of the QML system, such as access to certain components or intermediate data. This level of access enables more sophisticated attacks.
\begin{itemize}[label={--}]

    \item \textbf{Data Poisoning in Quantum Pipelines:} The adversary is assumed to operate within the quantum cloud infrastructure where a victim outsources the training of a hybrid quantum-classical model, having access to the victim's preprocessed and labeled training dataset and partial visibility into the QNN architecture, specifically, the classical-to-quantum data encoding circuit responsible for mapping classical inputs to quantum states in the Hilbert space (Fig. \ref{fig:poisoning}). However, the adversary does not possess knowledge of the PQC, the loss function, the optimizer, or other training dynamics. Within this constrained view, the adversary conducts an indiscriminate data poisoning attack by modifying the labels of a subset of the training data based on a quantum-specific criterion: intra-class encoder state similarity (ESS) \cite{kundu2025adversarialdatapoisoningattacks}. By computing the density matrices resulting from the encoding circuit and measuring quantum distances (e.g., Frobenius norm) between states, the adversary identifies and assigns incorrect labels that maximize intra-class dissimilarity. This manipulation deteriorates the structure of the encoded quantum feature space, thereby disrupting the training process and degrading the generalization capability of the QNN. Such a model degradation attack, achievable without full circuit transparency, exemplifies the threat posed by semi-privileged adversaries in practical quantum cloud environments where partial circuit disclosure and data access are common.

    \item \textbf{Vulnerabilities in Hybrid Quantum-Classical Architectures:} The adversary can also be modeled as an untrusted quantum cloud provider with partial but strategically significant visibility into the computation pipeline. Specifically, the adversary does not possess access to the original source code of the QML model, its training data, or the internal architecture design (e.g., parameter initialization, circuit topology, or training dynamics) \cite{wang2023qumosframeworkpreservingsecurity}. However, the attacker does receive access to the compiled, executable quantum circuits submitted for inference, which are composed of a classical-to-quantum data encoder $D(Z)$ and a parameterized quantum model $C(\theta)$. By impersonating a legitimate user and submitting chosen inputs $D(Z)$, the adversary induces the QML service to transmit the composed circuit $C(\theta) \cdot D(Z)$ to the cloud infrastructure. Leveraging the known structure of their own encoder $D(Z)$, the attacker can construct and apply the inverse circuit $D^{-1}(Z)$ to effectively cancel the data encoding portion, thereby isolating the underlying QML model $C(\theta)$. This exploit relies on the reversibility of quantum circuits and the transparency of the encoding function, allowing the adversary to reconstruct the core model even without access to its training history or high-level design intent. The attacker’s capabilities thus extend beyond a conventional black-box model, given their ability to observe and manipulate low-level quantum instructions. This gray-box attack illustrates a critical vulnerability in centralized QML deployments and motivates the need for distributed execution strategies, such as partitioning the QML model across multiple cloud providers to prevent any single adversary from recovering the complete model.

\end{itemize}

\begin{figure}[t]
    \centering
    \includegraphics[scale=1.0]{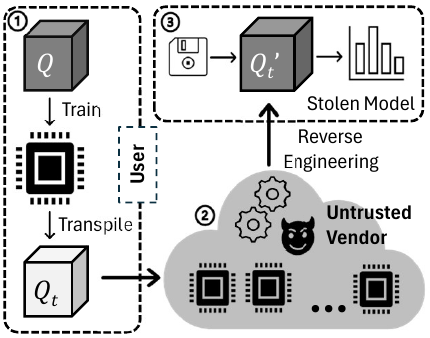}
    \caption{The flow diagram describes reverse engineering of QML parameters by untrusted third-party vendors acting as adversaries. (1) shows the user training and transpiling a QML model $Q$ using non-proprietary quantum hardware and sending the transpiled version of the trained model $Q_t$ to the untrusted vendor for inferencing. (2) and (3) describe the attack model involving the procedure of reverse engineering performed by the untrusted vendor to extract the parameters and steal the IP of the user-designed model.}
    \label{fig:reverse-engineering}       
\end{figure}

\subsubsection{White-Box Attack Models}
White-box adversaries have comprehensive access to the QML system, including its architecture, parameters, and training data. This access facilitates the most potent attacks.

\begin{itemize}[label={--}]
    \item \textbf{Circuit-Level Backdooring:} The adversary is modeled as a fully privileged, potentially malicious quantum cloud service provider that has comprehensive access to the transpiled, hardware-specific quantum circuit of a trained Quantum Machine Learning (QML) model at inference time. Although the adversary lacks access to the original high-level source code, training data, and optimizer trajectory, they possess the complete transpiled parameterized quantum circuit (PQC), which includes low-level gate sequences, qubit mappings, and optimized parameter placements. This level of transparency, typical in current QML-as-a-Service (QMLaaS) deployments, allows the attacker to reverse-engineer the architecture and parameter values of the original model \cite{ghosh2024quantumimitationgamereverse, ghosh2024aidrivenreverseengineeringqml} (Fig. \ref{fig:reverse-engineering}). The adversary reverse engineers the transpiled circuit by parsing it to extract the gate-level representation based on the backend’s basis gate set (e.g., $\{id, X, SX, CNOT, RZ\}$ in IBM’s superconducting platforms), the device’s coupling map, and transpilation optimization strategies. The attacker systematically analyzes the gate patterns, especially the sequences of $RZ, CNOT,$ and $SX$ gates, using a lookup table (LUT) constructed from known transpilation patterns of common parametric gates ($RX, RY, RZ$). For example, a transpiled pattern such as $RZ(\pi/2)\cdot SX \cdot RZ(\phi)\cdot SX \cdot RZ(\pi/2)$ can be mapped back to an $RX(\theta)$ rotation gate with an inferred parameter $\theta$ derived via decomposition. To recover the entanglement structure, the adversary reverses SWAP insertions and logical-to-physical qubit mappings using the device’s coupling constraints. For parameter recovery, a brute-force parameter sweep is conducted over reduced domains (e.g., $\theta \in [-\pi,\pi]$), leveraging access to the same transpiler toolchain and computational resources. The adversary iteratively generates reverse-engineered circuits, transpiles them, and compares their parameterized structure and output statistics against the golden reference circuit using a fidelity metric. The reverse engineering objective is to minimize the discrepancy between the reverse-engineered and original transpiled circuits, achieving functional equivalence. This capability enables the adversary to reconstruct an approximate but operational copy of the architecture-agnostic PQC, which can then be ported across different quantum backends, recompiled with alternative gate sets, or subjected to unauthorized modifications such as watermark tampering, adversarial re-training, or IP theft. The attack demonstrates that access to transpiled circuits alone—without training data or source code—is sufficient to compromise the confidentiality and deployability of QML models. 

    \item \textbf{Pulse-Level Attacks:} The adversary is assumed to have complete visibility into and control over the low-level representation of quantum circuits, specifically, the pulse schedules that encode gate operations for execution on real hardware. This threat model is grounded in a supply chain compromise scenario, where the attacker operates within the quantum software toolchain or an SDK (e.g., Qiskit, Amazon Braket), allowing them to embed malicious behavior into custom gate definitions without altering the abstract gate-level logic visible to the user \cite{xu2024securityattacksabusingpulselevel}. The adversary has access to the pulse-level attributes associated with each gate, including waveform shapes, amplitudes, durations, modulation frequencies, phase offsets, and control channel mappings. This enables them to craft tailored pulse-level attacks that are functionally correct at the logical level but deviate from the intended physical implementation. Two main classes of attacks are considered: (i) channel-level attacks, which manipulate the routing or scheduling of control pulses (e.g., qubit plundering by leaking operations across channels, blocking pulses to freeze victim qubits, or reordering pulses to disrupt computation), and (ii) pulse-level attacks, which directly alter signal parameters to degrade fidelity, introduce timing shifts, or induce cross-talk. Importantly, these attacks remain stealthy because pulse-level behavior is analog, dynamic (due to frequent calibrations), and largely opaque to users who verify circuits at the gate abstraction layer. The adversary exploits this abstraction gap, embedding payloads that evade detection even under standard circuit validation procedures. Furthermore, the attack model accounts for different classes of victims—ranging from naive users with no verification capability to sophisticated users who conduct unitary fidelity checks—by tailoring the pulse modifications to fall within hardware noise tolerances or mimic expected calibration drift. This comprehensive control over internal representations and execution pathways situates the attacker firmly in the white-box category and reveals a potent, underexplored threat vector in quantum computing: adversarial manipulation at the pulse control layer beneath the visible circuit abstraction.

    \item \textbf{Prompting the need for Quantum Logic Locking (QLL):} In cases where the adversary is characterized as an honest-but-curious quantum computing server—such as a commercial quantum cloud platform—or an internal malicious actor with full control over the quantum hardware and execution stack. The client, acting as a quantum circuit designer, compiles a high-level quantum program that includes a proprietary quantum oracle and submits the fully decomposed and technology-mapped physical quantum circuit to the server for execution. The adversary, by virtue of operating the hardware, has unrestricted access to all components of the physical circuit, including detailed gate-level descriptions, qubit mappings, scheduling metadata, and execution results \cite{10.1145/3358184}. This level of access enables the adversary to reverse engineer the structure and functionality of the embedded quantum oracle—often used in Grover’s search and other oracle-based algorithms—thus compromising intellectual property (IP) and sensitive problem-specific information encoded within the circuit. The attack model assumes that the quantum oracle constitutes proprietary logic encoded as a reversible sub-circuit and that its structure and behavior are critical to the confidentiality of the client’s computational objective. The adversary may observe measurement outcomes, simulate alternate input configurations, or perform structural analysis on the gate network to infer the function implemented by the oracle. To counter this, the proposed defense involves locking the oracle via the insertion of additional key-controlled quantum gates—most commonly CNOT gates controlled by dynamically reassigned key qubits—creating multiple functional modes of the oracle indistinguishable to the server. Only the client, who holds the secret key qubit schedule, can identify the correct functional output post-execution. This defense assumes the attacker can access and execute all oracle modes but lacks knowledge of the key qubit values that dictate the correct circuit semantics. As such, this model represents a fully privileged white-box adversary with both read and execution access, capable of structural analysis but unable to discern protected logic without cryptographic or obfuscatory barriers.

    \item \textbf{Obfuscating Phase in Quantum Circuit Compilation:} The adversary is modeled as an untrusted third-party quantum compiler with full access to the pre-compiled quantum circuit supplied by the circuit designer. This model assumes that the quantum circuit, in its algorithmic or logical form, must be transmitted to an external compiler for hardware-specific transpilation and optimization. In doing so, the designer inadvertently exposes the complete structure and functional logic of the circuit, including its gate sequence, entanglement topology, and qubit mappings. The adversary, operating at the compiler level, is thus capable of inspecting every component of the circuit, enabling the extraction of proprietary algorithms (intellectual property theft), the reverse engineering of core functionality, and the malicious alteration of gate logic to embed hardware Trojans or counterfeit modifications. This high-privilege access classifies the model squarely within the white-box paradigm, as the attacker does not need to infer information from indirect observations but instead directly sees and operates on the internal representations of the quantum circuit. The threat is exacerbated by the current reliance on widely available third-party compilers, such as Qiskit, TKET, and Cirq, which may be maintained by external vendors and integrated into diverse cloud-based quantum development environments. These compilers can modify phase relationships, alter gate structures, or introduce stealthy perturbations that are difficult to detect post-compilation \cite{rehman2025opaqueobfuscatingphasequantum}.
    
\end{itemize}

%% file: author/section3.tex
\section{Countermeasures against State-of-the-Art Attack Models}
\label{sec3}

The unique characteristics of QML systems necessitate specialized security measures to address their inherent vulnerabilities. Recent research by various scholars has led to the development of targeted countermeasures aimed at fortifying QML models against a spectrum of adversarial threats. This section delineates these strategies, emphasizing their applicability across different stages of the QML pipeline.

\subsection{Circuit-Level Obfuscation and Logic Locking}

Unlike classical logic locking where each key bit is implemented via dedicated key gates or control logic at the transistor level, QLL operates under fundamentally different constraints and opportunities. Quantum gates act on superposed and entangled states, and any direct mapping of classical key-bit-to-control structures is both physically expensive and computationally limiting. To address this, the E-LoQ \cite{10.1145/3358184} scheme introduces a key-embedded quantum locking framework, wherein multiple classical key bits are efficiently encoded into a single key qubit. This enables scalable obfuscation of quantum circuits such as those found in modern quantum machine learning models without incurring significant qubit or gate overheads.

The mechanism begins by transforming a quantum circuit $\text{Circ} = G_m \dots G_0$ into a locked variant $\text{Circ}'$, via a key-dependent encryption function $\text{Enc(Circ}, \vec{k})$, where $\vec{k} \in \{0,1\}^n$ is a classical key of length $n$. This transformation involves the addition of a dedicated key qubit $q_k$, which dynamically progresses through the $n$ key bits via interleaved Hadamard and Pauli-X operations, effectively implementing sequential key indexing within a single quantum wire.
Two forms of gate-level modifications are introduced in the circuit--
\textbf{Functional Gates:} A subset of original gates are replaced with key-controlled versions, where execution is conditional on the state of $q_k$ being $\ket{1}$. This ensures that the circuit performs correctly only when the applied key matches $\vec{k}$.
\textbf{Dummy Gates:} New gates are inserted whose operations are contingent on $q_k = \ket{0}$, effectively acting as no-ops under the correct key and injecting noise under incorrect ones.
The key scheduling across the circuit is randomized, and key bit transitions are obscured through ``H-masking," which adds Hadamard gates before each key-dependent operation to decorrelate adjacent key bits and resist simplification by transpilers. During the decryption phase $\text{Dec(Circ}', \vec{k})$, the correct sequence of Pauli-X and Hadamard operations is applied to reconstruct the original functionality, and the locked circuit is simplified by removing redundant identity gates and the auxiliary $q_k$.
Critically, the unlocked circuit $\text{Circ}''$ satisfies $\text{Circ}'' = \text{Circ}$ only if $\vec{k}$ is correct. When incorrect keys $\vec{k}_g \neq \vec{k}$ are used, both functional and dummy gates are misaligned, yielding a semantically corrupted circuit.
Security analysis of E-LoQ reveals its robustness against key guessing and reverse engineering. The obfuscation affects both the circuit’s functional output and structural layout, increasing the difficulty of reverse-engineering attacks. Quantitative metrics such as Total Variation Distance (TVD), Hamming Variation Distance (HVD), and Degree of Functional Corruption (DFC) consistently show significant deviation from the baseline circuit when incorrect keys are used—TVD and HVD often exceed 0.9, and DFC approaches -1 in benchmark evaluations.

Moreover, key guessing is empirically infeasible. The average success rate of randomly guessing the correct key is negligible, and even partial key matches yield circuits with incorrect outputs or high functional divergence. Experiments with various benchmark circuits, including arithmetic units, adders, and comparators, confirm that even short key lengths (e.g., 6 bits) provide sufficient entropy to prevent adversarial inversion, especially when keys are mapped to variable gate positions.
E-LoQ further ensures minimal overhead. Post-decryption simplification restores the original circuit depth with fidelity loss constrained below 1\%. Unlike earlier schemes that required one qubit per key bit, or retained permanent overhead in the deployed circuit, E-LoQ discards all auxiliary locking logic after compilation, rendering it highly practical for near-term QML circuits that operate under severe qubit and gate budget constraints.

\begin{figure}[t]
    \centering
    \includegraphics[scale=.8]{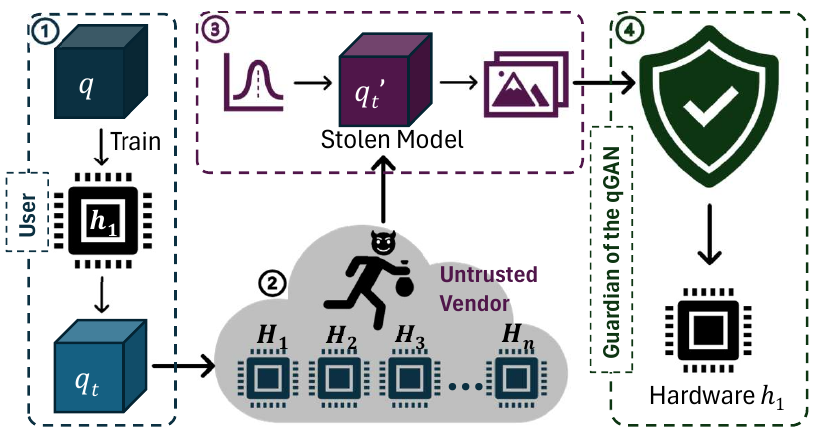}
    \caption{The flow diagram describes our attack model and the proposed security measure. In the figure (1) shows the user training his qGAN, $q$ on hardware $h_1$ to generate a trained qGAN $q_t$; (2), (3) describes the threat model of an untrusted quantum hardware vendor where the user sends qt for inferencing (note, the hardware used for inferencing, Hi, could be different than the hardware used for training $h_1$), from where it gets counterfeited by the untrusted vendor ($q_t^{`}$); (4) is our proposed method of collecting the images generated by $q_t^{`}$ and detecting the hardware where it has been trained using the classifier for proof of ownership.}
    \label{fig:qgan}       
\end{figure}

\subsection{Hardware-Aware Watermarking}

In the NSIQ era of quantum computing, noise plays an active role in the quantum circuit design and implementation. Leveraging this property of quantum hardware a passive, hardware-dependent watermark can be embedded in quantum neural networks. The authors in \cite{ghosh2024guardiansquantumgan} have demonstrated the watermarking scheme on quantum generative adversarial networks (qGANs). Unlike active watermarking strategies, this approach exploits the stochastic variations introduced during quantum circuit execution as a natural means of identifying the quantum backend used for training. These variations, being hardware-specific, are encoded into the trained model’s parameters and manifest in the output data generated by the qGAN (Fig. \ref{fig:qgan}).

During the training phase, the qGAN model is executed on a designated quantum backend. The model architecture consists of multiple sub-generators implemented as PQCs, which operate on latent vectors to produce image patches. Due to the sensitivity of PQCs to hardware noise (e.g., gate error rates, decoherence times, readout errors), the evolution of model parameters inherently reflects the unique noise profile of the training hardware. The generated images carry this noise imprint as a latent watermark, which remains intact even when the model is later inferred on different hardware.
To enable watermark extraction, a classical convolutional neural network (CNN) classifier is trained on images generated by qGANs trained on a suite of known hardware. The classifier learns to identify subtle statistical differences arising from the embedded noise, achieving near-perfect classification accuracy for models trained on single hardware and approximately 90\% for models trained across sequences of multiple hardware. For validation, the classifier’s output is filtered using a confidence threshold $M$; only predictions exceeding this threshold are accepted as proof of hardware provenance. This enables reliable differentiation between genuine and counterfeit models, even in adversarial scenarios involving unknown or tampered training sources.

Security analysis reveals several strengths of this watermarking strategy. First, the uniqueness of the watermark scales with the size of the available hardware suite. If a model is trained on a sequence of $k$ hardware systems from a set of $n$, the probability of collision reduces to $\prod_{i=1}^k \frac{1}{n - i}$, providing high uniqueness even for moderate values of $k$. For instance, selecting 5 backends from a pool of 15 yields a collision probability on the order of $10^{-5}$.
The watermark is resistant to removal because the training process is inherently non-reversible. Even if an adversary has access to the white-box architecture and transpiled quantum circuit, reconstructing the exact noise-induced parameter trajectory is computationally infeasible without knowledge of initial states and training data. Furthermore, attempts to tamper with the watermark—by fine-tuning the stolen model on new hardware—only partially overwrite the original noise signature unless extensive retraining is performed, which reduces the incentive to engage in such tampering.
False claims of ownership based on ghost watermarks are similarly deterred by requiring the classifier to match the watermark with high confidence and optionally through multi-hardware signature embedding. By training across an ordered sequence of backends, the watermark becomes both composite and sequence-dependent, making it extremely difficult to forge.
This approach introduces no architectural overhead, relies solely on existing training dynamics, and is broadly applicable to other QML models where noise-dependent parameter evolution can be captured and later authenticated. The method is particularly suited for cloud-based quantum service environments where trust boundaries are minimal and IP leakage risks are high.

\subsection{Side-Channel Attack Mitigation}
QML workloads, particularly those involving VQCs, are susceptible to crosstalk-induced side-channel attacks when deployed on multi-tenant NISQ hardware. The intrinsic physical interactions among proximate qubits can expose dynamic circuit features, including entangling gate patterns and scheduling information. To safeguard the structural confidentiality of QML circuits—whose layer depth, connectivity, and parameterized entangling operations are critical IP—several crosstalk-aware defense strategies have been developed. These aim to reduce the observability and reproducibility of physical noise signatures that side-channel adversaries may exploit.
\begin{itemize}[label={--}]
    \item \textbf{Qubit Layout Randomization and Dynamic Mapping: }To decouple logical qubit identities from their physical placement, a randomized mapping strategy is employed prior to execution. During transpilation, each QML model instance is compiled using a randomly selected allocation of logical-to-physical qubits, ensuring that sensitive entangling operations (e.g., $CNOT$ or $CR$ gates) do not consistently occupy fixed locations within the hardware topology \cite{choudhury2024crosstalkinducedchannelthreatsmultitenant}.
    This strategy is particularly effective for QML circuits where the placement of entangled qubit pairs conveys model structure, such as patch-wise variational layers in quantum convolutional neural networks or entanglement topologies in quantum graph learning models. By rotating the spatial context of each execution, the correlation between specific victim operations and observer noise signals is weakened, reducing side-channel signal coherence across executions.

    \item \textbf{Noise-Shaping: }A second line of defense introduces execution-time obfuscation through the insertion of padding operations and idle gate shaping \cite{xu2024securityattacksabusingpulselevel}. Selective insertion of single-qubit rotations and decoupling gates on both active and idle qubits during known sensitive intervals (e.g., high-density entangling layers) disrupts the crosstalk field's spectral consistency.
    For QML workloads, this is applied dynamically based on the learned temporal profile of the variational layers. For instance, entangling blocks that dominate the expressivity of the QML model (e.g., in quantum feature maps or variational classifiers) are targeted for padding with dynamically randomized gate sequences. These gates are selected to be unitary equivalent to identity in function, but sufficiently distinct in hardware-level control signals (e.g., shaped microwave pulse envelopes) to introduce temporal and spatial noise decoys.
    \item \textbf{Circuit Obfuscation: }Structural obfuscation of QML models at the transpilation layer is realized through compilation strategies that preserve functional equivalence while altering gate scheduling and topology \cite{choudhury2024crosstalkinducedchannelthreatsmultitenant}. Techniques include:
    \textbf{Gate Reordering:} Rewriting commutable gate blocks to alter execution timing without affecting output fidelity.
    \textbf{Dummy Gate Insertion:} Adding canceling $CNOT$ pairs or parameterized gates with zero-angle rotations to obscure true operation density.
    \textbf{Template Substitution:} Replacing standard subcircuits with functionally equivalent but structurally distinct alternatives.
    These methods preserve the training trajectory and inference behavior of the QML model while masking structural invariants that may be leveraged by adversaries to perform classification or reconstruction attacks on variational models.

    \item \textbf{Randomizing the Execution Scheduling: }To mitigate timing-based leakage exploited through temporal bucketing analysis, a stochastic scheduler introduces non-determinism in circuit queuing and execution windows \cite{xu2024securityattacksabusingpulselevel}. Each QML job is fragmented and re-ordered with randomized delays or submitted as batched sequences interleaved with decoy circuits. The scheduler may leverage dummy jobs from a secure controller to increase entropy in system-wide temporal load patterns.
    This is particularly relevant to QML inference-as-a-service deployments, where repeated invocations of a model on user data must maintain indistinguishability at the execution profile level. Even if the model’s layout is consistent, variance in gate timing and circuit duration limits adversarial correlation of output signal deviations to specific circuit epochs.

    \item \textbf{Topology-Aware Device Assignment and Isolation: }Lastly, QML security-aware runtime systems may employ physical isolation by assigning sensitive model components to minimally coupled regions of the quantum processor. Device regions with reduced crosstalk coefficients are selected for model qubits, while known high-coupling zones are reserved for untrusted or third-party tenants \cite{choudhury2024crosstalkinducedchannelthreatsmultitenant}\cite{xu2024securityattacksabusingpulselevel}.
    In hybrid QML models where classical and quantum inference are pipelined, this approach allows for staging of the quantum kernel execution into pre-designated hardware zones, minimizing the physical leakage paths accessible to concurrent adversaries.
    
\end{itemize}

\subsection{Secure Model Partitioning and Distribution}

To mitigate model theft risks associated with untrusted quantum cloud providers, the QuMoS framework \cite{wang2023qumosframeworkpreservingsecurity} introduces a partitioning-based security strategy for QML workloads. Rather than relying on cryptographic obfuscation or post-training watermarking, QuMoS decomposes the quantum model into a set of computational submodules (nodes), each of which is executed on a different, physically isolated quantum backend. This architectural fragmentation ensures that no single provider has access to the entire quantum circuit, thereby preventing reconstruction of the full QML model from any individual subsystem.
Each submodule of the QML model is defined as a quantum computational block $C_i(\theta_i)$, representing a distinct portion of the variational quantum circuit. The blocks are linked via a directed acyclic graph (DAG), with edges denoting data flow and inter-node dependencies. The QuMoS runtime dispatches these blocks for execution on separate quantum cloud providers, with connectivity managed through secure classical post-processing of intermediate measurement results.
Security stems from the design constraint that any single submodel (i.e., connected subset of blocks assigned to a single provider) must exhibit low standalone predictive power. That is, no subset executable on a compromised provider should retain a nontrivial approximation of the full model's functionality. To enforce this, the framework defines and optimizes a heuristic metric $\text{SecMec}(M)$, computed as:
\[
\text{SecMec}(M) = 1 - \max_{q \in \mathcal{Q}_S,\, sm \in \mathcal{S}_M} \frac{\text{ACC}(sm, q)}{\text{ACC}(M)}
\]
where $\mathcal{Q}_S$ is the set of quantum devices, $\mathcal{S}_M$ is the set of security submodels (i.e., all maximal subgraphs confined to one provider), and $\text{ACC}$ denotes accuracy on a test dataset. This metric quantifies how closely any submodel can approximate the full QML behavior; higher values indicate greater resilience to partial model extraction.
To identify optimal partitionings, QuMoS employs a reinforcement learning-based controller. The controller explores the discrete space of node-to-node topologies, quantum architecture mappings, and provider assignments. Each candidate model is evaluated on the basis of its accuracy and $\text{SecMec}(M)$, and controller parameters are updated using a reward signal $R \propto \text{ACC}(M) + \lambda \cdot \text{SecMec}(M)$, where $\lambda$ modulates the trade-off between performance and security.

Each candidate architecture is trained and validated across a suite of simulated noisy backends (e.g., IBMQ quito, belem, manila), with security submodels identified through topological analysis of the deployment graph. Submodels are then separately evaluated for inference performance on their assigned provider to calculate their leakage potential.
Empirical evaluation shows that naive partitioning schemes, even when distributed across multiple providers, often fail to secure the model: some submodels retain disproportionately high accuracy, enabling partial model theft. In contrast, architectures optimized by the QuMoS engine consistently maximize security by ensuring that submodels are entangled in non-trivial data dependencies, reducing their standalone efficacy. Furthermore, QuMoS discovers secure model configurations with minimal loss in accuracy—often within 1–2\% of centralized neural architecture search (NAS) baselines, and occasionally outperforming them.
Notably, the framework is robust to varying numbers of available providers. Even with only two providers, it can achieve substantial security benefits, and the partitioning strategy dynamically adapts to available infrastructure. Moreover, the system selectively omits unnecessary nodes and providers to optimize the balance between model fidelity and confidentiality.

\subsection{Hardware-Induced Output Perturbation}

To mitigate threats pertaining to the cloning of QML models by repeated querying, authors in  \cite{kundu2024evaluating} introduce a perturbation-based defense mechanism that leverages the intrinsic noise characteristics and hardware heterogeneity of NISQ systems to obfuscate output distributions and degrade the fidelity of cloned QML models.
Two concrete techniques are proposed: \textbf{Hardware Variation-Induced Perturbation (HVIP)} and \textbf{Hardware and Architecture Variation-Induced Perturbation (HAVIP)}. Both exploit naturally occurring device-level variances and architectural diversity across quantum backends to introduce controlled but unpredictable distortions in the QNN outputs.

\begin{itemize}[label={--}]
    \item \textbf{HVIP: }HVIP operates by dynamically varying the execution backend of the cloud-hosted QNN. Rather than consistently executing all inference queries on a single quantum device, the system randomly routes each incoming query to a different hardware instance. Since quantum backends differ significantly in their coupling maps, gate fidelities, basis gate sets, and noise profiles (e.g., amplitude damping, depolarization, readout error), the resulting measurement statistics diverge subtly across devices.
    This variability introduces stochastic perturbations into the output probability vectors returned to the adversary, even when the same input is queried repeatedly. As a result, the attacker accumulates a training dataset with inconsistent label distributions, thereby impairing the convergence and fidelity of the substitute QNN. HVIP is especially effective when the backend pool includes devices with non-overlapping error characteristics, amplifying divergence in the softmax output distributions.
    \item \textbf{HAVIP: }HAVIP enhances the defense by jointly introducing architectural heterogeneity into the system. Instead of deploying a single QNN across multiple devices, HAVIP trains multiple, distinct QNN architectures—each optimized and compiled for a different quantum backend. Upon receiving a query, the inference engine selects one of these QNN instances at random to process the input.
    Since each model is structurally distinct (e.g., differing in circuit depth, entanglement structure, parameter count), and each backend exhibits unique error characteristics, the ensemble produces non-deterministic, architecture-specific outputs. This misaligns the decision boundaries perceived by the adversary and fragments the label distribution in the adversarial dataset.
    The combination of circuit-level and hardware-level diversity ensures that the attacker is unable to consistently map input features to reliable output labels, leading to noisy, low-fidelity clones even when high-volume queries are issued.
\end{itemize}
The defense mechanisms were tested on hybrid QML pipelines involving parameterized quantum circuits (PQC-1, PQC-6, PQC-17, PQC-19), using MNIST, Fashion-MNIST, Kuzushiji, and Letters datasets. Clone models trained on Top-1 and Top-k label queries were evaluated across scenarios with and without defense activation.
Quantitative results show that HAVIP and HVIP introduce measurable perturbations: up to 15.71\% label mismatches in Top-1 queries and 10.2\% TVD in Top-k outputs. The cloned models experienced performance drops of up to 13\% in test accuracy, particularly when trained with probability vectors returned from the perturbed ensemble.
Interestingly, the study also observes that QML models trained in noisy environments (e.g., using SPSA optimizers and mixed-device noise simulations) exhibit higher baseline robustness to minor perturbations. This dual-edged outcome implies that while HVIP and HAVIP degrade attacker success, they must be reinforced with more aggressive or adaptive noise modulation strategies to remain effective as quantum devices scale and standardize.

From a QMLaaS perspective, HVIP and HAVIP represent practical and cost-effective methods to enhance IP protection. These defenses do not require cryptographic encapsulation or quantum authentication but rather exploit the inherent stochasticity of NISQ systems. By treating noise and architectural diversity as security assets rather than liabilities, the system obfuscates functional mappings without compromising user-level inference performance, aligning well with the QML workloads, particularly where small decision boundary shifts or softmax noise can drastically affect training convergence in adversarial cloning attempts.

%% file: author/section4.tex
 \section{Designing Secure QML Systems}
\label{sec4}

As QML transitions from theory to implementation, ensuring the security of these systems from the ground up is no longer optional—it is essential. Given the hybrid architecture of most QML systems, the high cost of training, and the increasing reliance on cloud-based quantum services, model developers must proactively incorporate security principles throughout the design lifecycle. This section outlines key guidelines for designing QML models with robustness and trustworthiness in mind, informed by state-of-the-art countermeasures and known attack vectors.

\subsection{Identifying Threat Model}

The development of a secure QML pipeline must commence with the formalization of a threat model that characterizes potential adversarial capabilities and objectives. In the context of QML systems, adversaries can be classified by their level of access: black-box, gray-box, or white-box. Black-box adversaries, typically external clients interacting through public QMLaaS APIs, are limited to input-output queries and rely on statistical probing to infer model behavior. Gray-box adversaries may be semi-privileged actors such as cloud providers or software vendors who have partial visibility into compiled circuits, encoders, or training data. White-box adversaries represent the most powerful threat, with full access to model internals, circuit parameters, transpiled gates, and even low-level pulse schedules. Defining the access level and intent—whether model extraction \cite{ghosh2024guardiansquantumgan}, data poisoning \cite{kundu2025adversarialdatapoisoningattacks}, or logic subversion \cite{farhi2018classificationquantumneuralnetworks}—is essential to selecting appropriate defenses and embedding security constraints throughout the QML pipeline.

\subsection{Data Encoding Security} 
The data encoding stage, which transforms classical inputs into quantum states, is inherently vulnerable due to its exposure at the initial point of interaction. Encoding strategies, such as amplitude or angle encoding, can leak structural information if left unprotected. To mitigate these vulnerabilities, encoding procedures must be designed to introduce entropy and obfuscation, ensuring that the mapping from input space to Hilbert space is not trivially invertible. Encoders can employ randomized gate structures or parameterized transformations whose specifics are concealed from execution environments. Furthermore, encoded data must undergo quantum-aware validation to prevent adversarial manipulation, particularly label poisoning attacks that exploit the similarity structure of quantum states. Incorporating noise-tolerant and adversarially robust encoding frameworks can prevent perturbation-induced degradation of quantum feature spaces, thus preserving the semantic integrity of the learning process.

\subsection{Obfuscate Quantum Circuit Structure}
Variational quantum circuits, central to QML models, are a primary target for model extraction and reverse engineering attacks due to their rich representational capacity and high design cost. The structural protection of such circuits requires a combination of logic obfuscation and conditional execution gating. Techniques such as QLL introduce key-dependent behavior by embedding control logic into circuit pathways, ensuring correct output generation only when a secret key is applied. An effective implementation, such as the E-LoQ scheme \cite{10.1145/3358184}, leverages a single qubit to sequentially encode multiple classical key bits, significantly minimizing overhead while injecting high entropy into the circuit behavior. Additionally, structural diversification through template substitutions, insertion of canceling gate pairs, and reordering of commutable blocks hinders attempts to statically analyze or simulate the original functional logic. These measures are vital in protecting the intellectual property encoded within the circuit and deterring adversarial replication.

\subsection{Harden the Hybrid Training Loop}

The quantum-classical interface in QML training pipelines constitutes a critical attack surface. Variational algorithms rely on classical optimizers to iteratively update quantum parameters based on measured outcomes, creating a bidirectional data flow that can be intercepted or altered. To secure this loop, communication between quantum and classical components must be authenticated and, where feasible, encrypted. Sensitive artifacts such as gradient values or cost functions must be safeguarded against both passive eavesdropping and active manipulation. Furthermore, robust training strategies can be deployed to preemptively inoculate models against poisoning and convergence manipulation. Adversarial training that incorporates quantum-aware perturbations during training, such as those induced by the QUID framework \cite{kundu2025adversarialdatapoisoningattacks}, ensures that the model's decision boundaries remain resilient to encoded state-space anomalies. Extensions of differential privacy to quantum optimization may further obfuscate parameter evolution trajectories, mitigating inference risks associated with repeated observations of intermediate states.

\subsection{Integrate Hardware-Aware Defenses}
Quantum hardware, particularly in the NISQ regime, exhibits idiosyncratic noise profiles and physical crosstalk that introduce side channels exploitable by adversaries. To defend against such leakage, hardware-aware mitigations must be integrated at the transpilation and scheduling layers. Logical-to-physical qubit mappings should be randomized across executions to prevent spatial correlation of entangling gates. Gate-level noise shaping, through the insertion of pulse-equivalent identity operations, can disrupt spectral signatures that would otherwise reveal circuit timing or structure. Furthermore, the stochastic fragmentation of execution schedules introduces temporal entropy, preventing timing analysis that correlates inputs with execution duration or gate activity. A complementary approach leverages the inherent noise fingerprint of quantum hardware as a passive watermark, uniquely tying a trained model to the backend used for its development. Such watermarks, when embedded within parameter evolution during training, are non-replicable without access to the original hardware and can be detected through trained forensic classifiers, providing a reliable means of provenance verification.

\subsection{Partitioning and Distributing Securely}

Given the risks associated with centralized deployment, particularly in cloud-hosted environments, secure QML systems must consider partitioning strategies that distribute model functionality across disjoint execution domains. The QuMoS framework exemplifies this approach by decomposing the quantum model into subcircuits, each executed on a distinct quantum backend. These submodules are designed to be individually non-functional, i.e., their inference accuracy is statistically indistinguishable from random guessing, thereby preventing adversaries from reconstructing meaningful model behavior even in the event of a breach. The orchestration of inter-submodule communication is handled through secure classical channels, often with post-measurement aggregation and normalization. Optimal partitioning is guided by reinforcement learning algorithms that balance task accuracy with the submodels’ predictive leakage, quantified via metrics such as $SecMec$. This architectural isolation paradigm significantly raises the difficulty threshold for model reconstruction and enables compliance with data locality and trust boundary requirements in federated quantum computing scenarios.

\section{Future Directions in Secure QML Design}
Future work in secure QML should systematically expand the threat landscape to encompass diverse attack vectors that exploit the unique properties of quantum computation. While current models address threats such as model stealing, data poisoning, and circuit backdooring, more sophisticated adversaries may leverage quantum-specific features, such as entanglement structure, noise profiles, and circuit reversibility, to execute covert or multi-stage attacks. For instance, the potential for adversarial manipulation through entanglement injection, inference-time noise pattern analysis, or optimizer-level gradient hijacking in hybrid setups remains largely unexamined. Additionally, as QML systems increasingly adopt QMLaaS and multi-tenant execution environments, new avenues for cross-user leakage and supply chain compromise need to be analyzed thoroughly for security gaps.

On the defensive front, advancing beyond isolated protections toward integrated and adaptive security frameworks is critical. Existing techniques, such as quantum logic locking, hardware-aware watermarking, and circuit obfuscation, have demonstrated effectiveness but are typically static and context-specific. Future defenses should be dynamic, incorporating runtime-aware scheduling, reconfigurable circuit topologies, and adversarially robust training regimes that reflect quantum state-space geometry. The idea of differential privacy also holds promise in the QML landscape for reducing leakage and enhancing confidentiality in untrusted environments. Importantly, the field would benefit from standardized evaluation protocols, threat benchmarking tools, and formal guarantees tailored to quantum computational settings.